\documentclass[conference]{IEEEtran}
\IEEEoverridecommandlockouts
\usepackage{cite}
\usepackage{amsmath,amssymb,amsfonts}
\usepackage{algorithmic}
\usepackage{graphicx}
\usepackage{textcomp}
\usepackage{xcolor}

\usepackage{eurosym}
\def\BibTeX{{\rm B\kern-.05em{\sc i\kern-.025em b}\kern-.08em
    T\kern-.1667em\lower.7ex\hbox{E}\kern-.125emX}}
\begin{document}

\title{Air Quality Control through Bike Sharing Fleets
\thanks{This work is partially funded by the Department of Mathematics of the University of Padua through the BIRD191227 project.}
}


\author{\IEEEauthorblockN{Stefan Ciprian Voinea, Armir Bujari, Claudio Enrico Palazzi}
\IEEEauthorblockA{\textit{Department Mathematics} \\
\textit{University of Padua}\\
	Padua, Italy \\
	stefanciprian.voinea@studenti.unipd.it, abujari@math.unipd.it, cpalazzi@math.unipd.it}
}

\IEEEoverridecommandlockouts
\IEEEpubid{\makebox[\columnwidth]{978-1-7281-8086-1/20/\$31.00~\copyright2020 IEEE \hfill} \hspace{\columnsep}\makebox[\columnwidth]{ }}

\maketitle

\IEEEpubidadjcol

\begin{abstract}
Air quality and the presence of tiny particular matter are crucial factors in human health, especially when considering urban scenarios. In this context, smart mobility coupled with low-cost sensors can create a distributed and sustainable platform for social sensing able to provide pervasive data to citizens and public administrations. Sustainable and eco-aware decisions can then be supported by empirical evidence, resulting in an improved life and city administration. In this paper, we present ArduECO, a simple Arduino-based wireless device able of collecting air quality data. Without loss of generality, we have designed our device as a box that can be installed on a bike; in this way, beyond private bikes, municipalities could exploit their bike sharing fleets as pervasive sensing systems.
\end{abstract}
%


\section{Introduction}\label{intro}
People and governments around the world are more and more aware of the importance of air quality and how much having clean and pollution-free air can influence our lives, both on the small and large scale, especially in urban scenarios\cite{delnevo19,mirri1,prandi17}.
To have clean air, all of us have to do something to avoid polluting it.

For short commutes, bicycles, skateboards and scooters have started to arise and conquer cities, either in their battery-powered or old-school human-powered form.
Since not everyone necessarily owns one of these green methods of transportation,
various companies have seen the opportunity to enter the market of shared transportation, proposing a \textit{pay-per-use} solution for bikes, electric scooters and other similar vehicles.	
Each company has its own phone application, network infrastructure and smart devices aboard their vehicles, gathering data like GPS (Global Positioning System) position, time spent by the user, parking spot, etc. to send them in the cloud and compute information like the cost of the ride and charging it to the customer.

All this falls under the \textit{IoT} (Internet of Things) paradigm, which has become a well-described market with new ideas and business opportunities being presented every day\cite{IoV18,Furini18,fanet17}.
Among the data collected by these companies, none is about air pollution.

This article describes \textit{ArduECO}, a wireless device based on an Arduino-like board capable of gathering data about air quality (and more with simple extensions) and sending them to the cloud, to be processed and displayed.
To this aim, taking inspiration from related work and without loss of generality, we have designed our tool as a device that can be attached to a bicycle\cite{Liu15,canarin}.
Although any citizen could install ArduECO on her/his bike, we believe that municipalities should play a major role in this by endowing their bike sharing fleets with it in order to create an effective pervasive sensing systems.


\section{Problem Statement and Background Overview}\label{bck}
Tracking air quality is an important task usually tied to weather agencies involving trucks seen parked or going around the city, collecting data like \textit{PM 2.5} and \textit{PM 10} values.
\textit{PM} stands for \textit{Particulate Matter} and indicates the term for a mixture of solid particles and liquid droplets found in the air, with diameters around 2.5 or 10 micrometers.
Some particles, such as dust, dirt, soot, or smoke, are large or dark enough to be seen with the naked eye, others can only be detected using an electron microscope.

In many cities, a great percentage of PM production has been cut down by decreasing the number of cars that can circulate and introducing more ways of public transportation: for example pay-per-use bike services have seen in the last year a vast amount of users.
For instance, since Mobike's launch in Padua, Italy, in 2019, its users have biked over 330,000 Km across the city and reduced CO{\footnotesize 2} emissions by over 60 tons.
CO (Carbon Monoxide) and CO{\footnotesize 2} (Carbon Dioxide) are gasses that become important pollutants when associated with cars, planes, power plants, and other human activities that involve the burning of fossil fuels such as gasoline and natural gas.

Two general strategies can be used to deploy an environmental monitoring system: creating a network of fixed sensors and resorting to mobile sensors. The former can monitor only predetermined locations where the sensors are deployed\cite{Moltchanov15}. The latter allows to monitor different parts of the city with the continuous and pervasive movements of the sensors, exploiting mobile units, smartphones and other wearables \cite{Alvear16,Al-Ali10,PathS}.

Focusing on low-cost mobile sensing units, Cambridge University proposed hand-held devices with electro-chemical sensors to collect and store data or send them in real time to a central computer for analysis\cite{cambridge}.
Similar devices exist also as proprietary devices sold
to consumers interested in their personal well-being.
These devices are usually implemented using low-cost sensors like the ones in MQ-series, made of a heating element an electro-chemical sensor, which will react only when it reaches certain temperatures.
ArduECO has been designed using the \textit{MQ-7} sensor, sensible for \textit{Carbon Monoxide}.

\section{Proposed solution}\label{solution}
	
\begin{figure}[t]
			\centerline{\includegraphics[width=8.8cm]{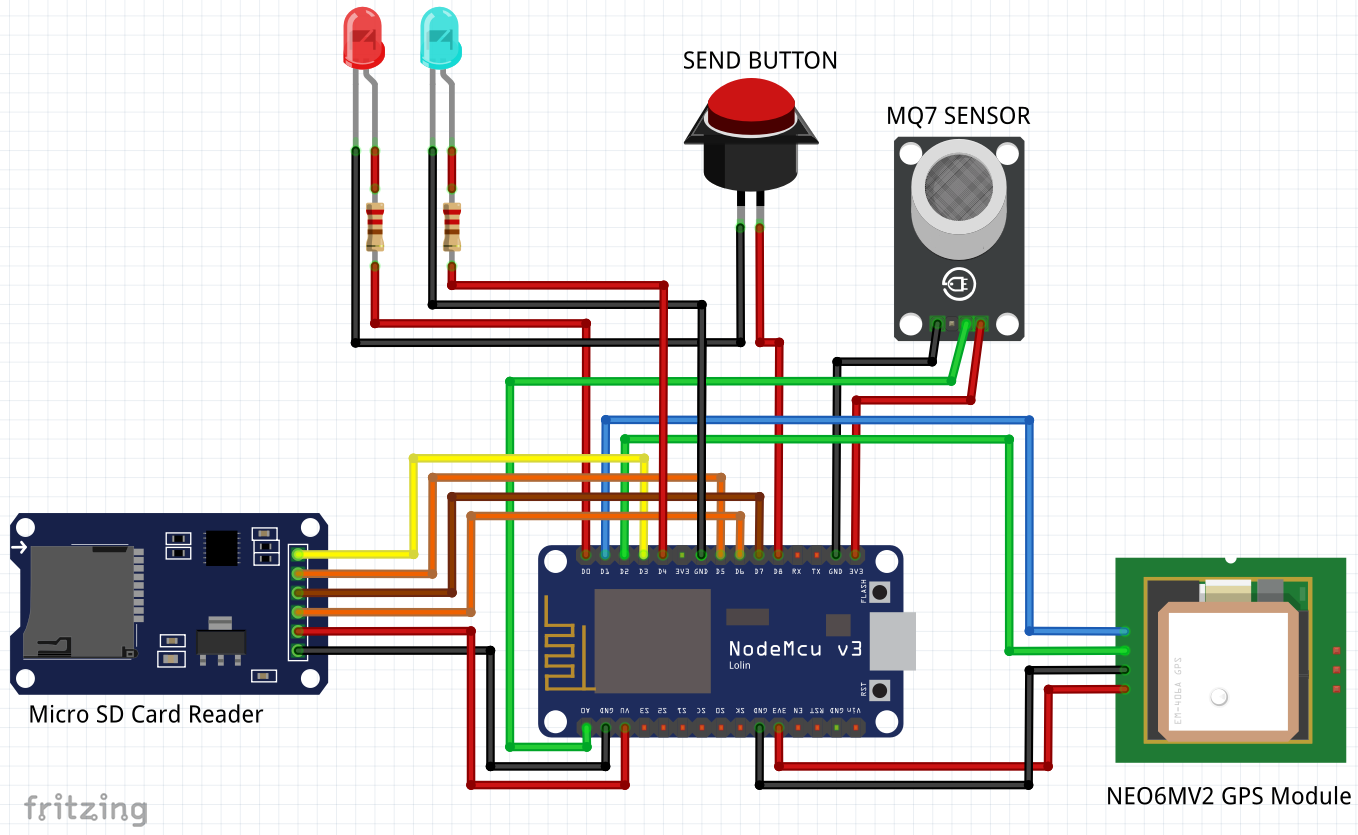}}
			\caption{ArduECO schematics}
			\label{schematics}
		\end{figure}

\begin{figure}[t]
			\centerline{\includegraphics[width=5.8cm]{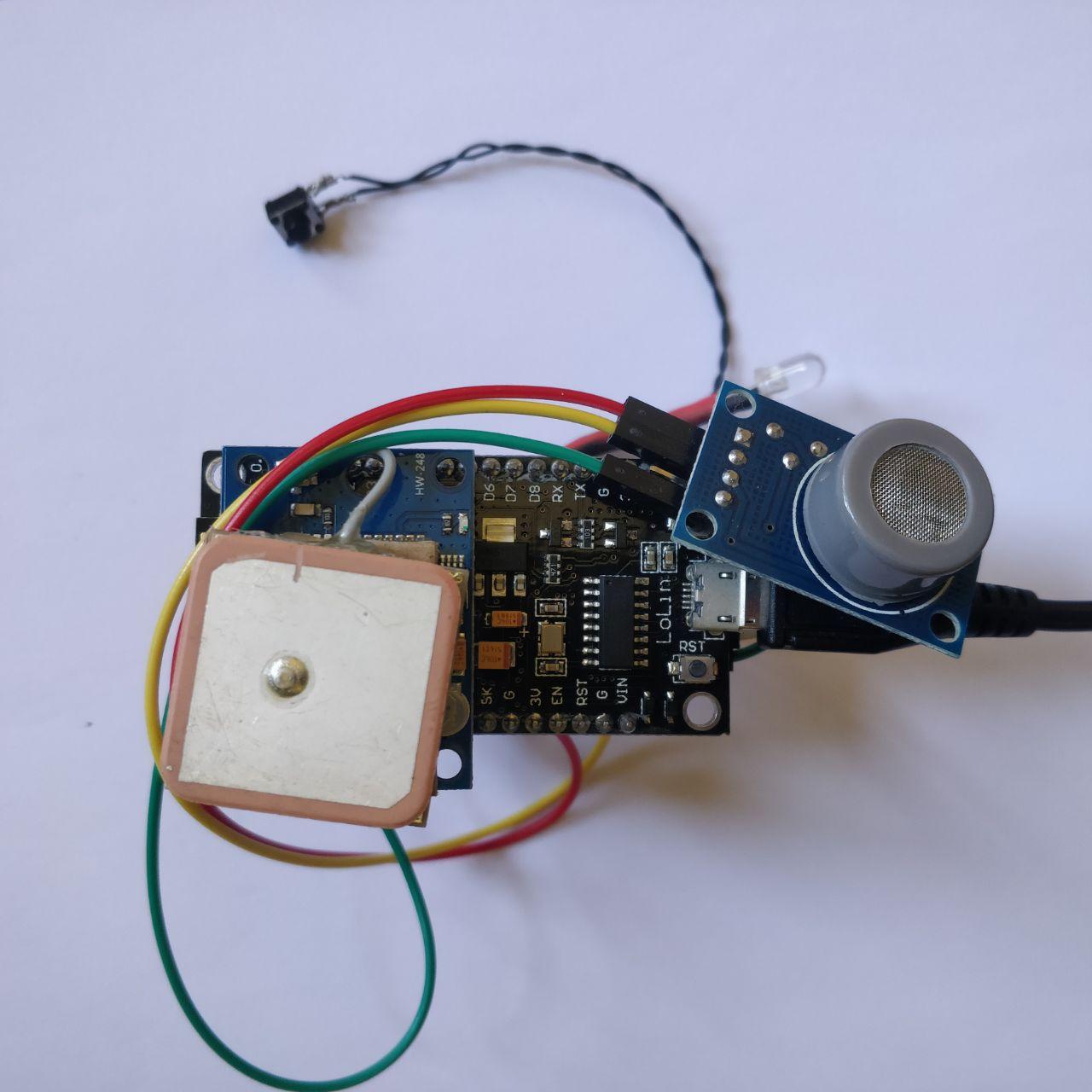}}
			\caption{ArduECO device}
			\label{device}
		\end{figure}

ArduECO is a wireless IoT device that represents one of the possible combinations of open source hardware and software.
This wireless device is able to detect the amount of CO and the level of pollution in the air in order to send this information to the cloud over Wi-Fi.
The schematics for this project, designed using the Fritzing CAD software for the design of electronics hardware and can be seen in Fig.~\ref{schematics}, whereas the real device is shown in Fig.~\ref{device}.


The goal is to design and create a device small enough to fit on several means of transportation, including our chosen case study, i.e., bicycles.
ArduECO represents indeed a viable solution: for instance, it could easily fit under the basket and be powered by batteries, solar cells or dynamos that already power the headlight and other circuitry allowing this bike sharing fleets to be connected with the vendor's cloud.
Given the modularity of ArduECO, its circuit can easily be adapted to fit other cases and forms.

\subsection{The circuit}\label{circuit}
	
		ArduECO is composed by four main components:
		\begin{itemize}
			\item \textit{NodeMCU}: this board embodies the heart and brain of the device as an an open-source development kit based on the ESP8266 chip that allows for prototyping IoT devices;
			\item \textit{MicroSD card reader}: it allows for collecting and keeping data cached for the current ride and permanently on a removable memory card;
			\item \textit{GPS sensor}: this is used to localize the device and has an onboard LED that blinks when it has established communication with a satellite;
			\item \textit{MQ sensor}: the MQ-7can easily be exchanged with other sensors that use the same board and connectors.
		\end{itemize}

The other components, LEDs and button, are used as I/O for interacting with the user.

In Section \ref{improvements}, we explain how the project can be improved both in hardware and software.

		In Fig.~\ref{schematics} there is no indication of a power supply since this prototype version has been powered using an external battery with a standard output of 5V and 1A.

\subsection{The Arduino software}
	
For ArduECO we chose to use Arduino's programming language instead of MicroPython because of the large community supporting it and the vast interoperability with other boards.
The main functions in the Arduino programming language are ``\textit{void setup()[]}'' and ``\textit{void loop()[]}'': the first one runs a single time when the device boots initializing and setting the initial values while the second one loops consecutively, allowing the program to change and respond.
ArduECO.ino is the main file executed by the NodeMCU and contains the previously described functions.

In ArduECO's software, the setup function contains the instructions to correctly initialize serial communication with the GPS module, create the files in the SD card, initialize the pin-outs, gather the certificates for communicating with the cloud and create a random id for identifying the ride, since each time the device boots is considered as a single ride.

An important part of the configuration of ArduECO is the file \textit{params.json}.
This file contains the configuration of several parameters such as the network SSID and password, topics for communicating with the MQTT server and its endpoint (network address).
If this file is not present in the SD card at boot-time and correctly formatted, ArduECO will not end the boot sequence, returning an error and rebooting after 10~s.

The loop function contains the instructions for the board to acquire a reading of its surroundings every 5~s by first detecting if a GPS fix if present and by reading the value from the analog pin connected to the MQ sensor.
Each read is then appended on two separate text files (\textit{cache\_log.txt} and \textit{perm\_log.txt}) in JSON format.
The \textit{cache\_log.txt} file is created each time ArduECO boots is related to a single session while \textit{perm\_log.txt} contains reads about each ride taken, including the sensor's id, the air quality and the coordinates of the data.

The external LEDs are used to communicate when the setup sequence has finished correctly or with errors, as well as for indicating whether the SSID specified in the \textit{params.json} file is in range and ArduECO is transmitting data.
When the button is pressed, ArduECO searches the list of networks in range for the specified one and, if present, connects to it.
Afterwards, it establishes communications with the MQTT server specified by the endpoint, also sending an initial message notifying the number of the following messages that will be sent.
Each row of the cache file is then transmitted to the server.
At the end of this sequence the cache file is deleted and created again awaiting new reads.

The other files that implement this project contain specific functions for either setting up the device or connecting the server.

\subsection{The cloud}
	
When the button is pressed, ArduECO connects to the access point with the SSID specified in the \textit{params.json} file and sends the contents of \textit{cache\_log.txt} to the cloud, specifically to AWS (Amazon Web Services).
\textit{IoT core} and \textit{Lambda} are the services that are used for this project.

ArduECO sends data to the MQTT server in IoT core specified as \textit{endpoint} in \textit{params.js}.
It connects to this server using certificates created in the IoT core console and that are loaded in ArduECO's memory, this means that they are uploaded via the Arduino IDE.
The MQTT (Message Queuing Telemetry Transport) server is designed as a lightweight messaging protocol that uses publish/subscribe operations to exchange data between clients and server, fast in data transmission.

\begin{figure}[t]
			\centerline{\includegraphics[width=8.8cm]{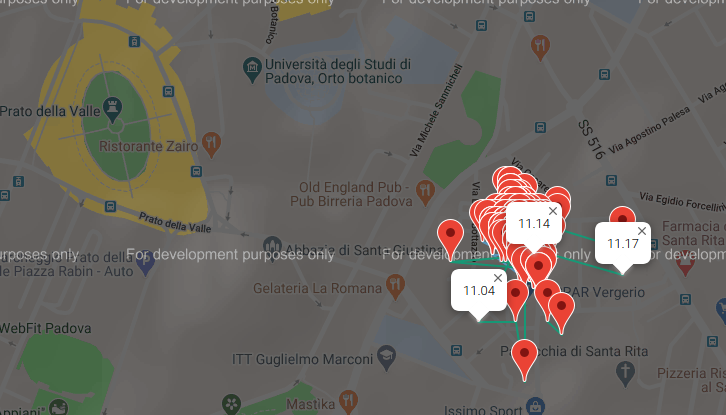}}
			\caption{Frontend map with markers and reads from ArduECO.}
			\label{altervista}
		\end{figure}

This service in IoT core is connected to a serverless Lambda function written in Python, triggered each time a new message arrives from ArduECO, which receives the JSON message as a parameter and calls a PHP page on
ardueco.altervista.org, where the front end of the service is hosted.
Altervista is a free website hosting platform with PHP and MySQL database.
After being saved in the MySQL database in Altervista is then displayed on the homepage using Google Maps' APIs.
Therefore gathered, aggregated and elaborated data can be displayed on a frontend map, with each marker indicating a reading from the device, as shown in Fig.~\ref{altervista}.
Each marker on the map represents a reading and, if clicked on, shows the value read by ArduECO's sensor.
Markers are connected with different colors indicating the ride they belong to.

\subsection{Hardware cost}
	
Because the components used for this project are open-source, they can be found sold on various websites from many vendors.
Here are some examples of prices taken from the Chinese e-commerce AliExpress at the time of writing.

		\begin{itemize}
			\item NodeMCU: $ \sim $ 2\euro
			\item MicroSD car reader: $ \sim $ 0.50\euro
			\item GPS sensor: $ \sim $ 3.5\euro
			\item MQ gas sensor: $ \sim $ 1\euro
		\end{itemize}


The total cost of these items amounts to $\sim$ 7\euro, considering shipping and costs of remaining components (cables, LEDs, button, SD card), it can arrive to less than 15\euro.
Note that the indicated prices refer to single pieces and in case of bulk purchases (at least 5 or 10 of each piece) the total cost would be much lower.
	
%
%
%

\section{Future improvements}\label{improvements}
 ArduECO is a feasible implementation that the considered hardware and software can create; yet, there are various improvements which could bring to an even better device.
	
\subsection{Hardware improvements}
The circuit described in Section \ref{circuit} has a single air sensor due to the limitation of the NodeMCU and ESP8266 chip that have only one analog input pin.
By connecting the board to an \textit{ADC} (Analog to Digital Converter) it becomes possible to add multiple sensors expanding the analog inputs.
ADCs convert an analog voltage on a pin to a digital number.

In order to upgrade the transmission method and use mobile data instead of Wi-Fi (which is limited to 2.4GHz on the NodeMCU) it is possible to switch the board to an Arduino Leonardo or an Arduino Uno.
Modules capable to add mobile connectivity can be easily found on sites like AliExpress
 and do not brig much additional overall cost (in this case $ \sim $ 2\euro).

Given the small size of the components, this project can be easily fitted within a custom made box easily installable on a bicycle.
A solution could be to design and 3D print a custom plastic case in which fit the whole device, considering that the air sensor should remain outside of the case and the whole device could be exposed to high levels of humidity especially during the winter season.

Powering is one important issue for IoT devices.
The prototype described in this manuscript runs on an external battery, but it can be coupled with other energy sources like a solar cell or a dynamo, so that it can be powered by movement of the wheel.
	
\subsection{Software improvements}	
A control could be added to check the reception of messages by the IoT server and, in case of a timeout or an error, the message could be sent again.

In this project various libraries have been used for the Arduino sensors, an improvement would be to better adapt these libraries for ArduECO, especially the MQ-7 library.

Furthermore, since ArduECO is an IoT device and battery is an essential need, adding a deep sleep mode could improve the energy performance.

\subsection{Cloud services improvements}
AWS' IoT core could be configured to have more MQTT servers based on the groups of devices divided by each city or zone in which they're used.

Another significant change would be using an AWS EC2 instance to host the website instead of Altervista.
Considering an architecture that could easily scale horizontally, and supporting more devices connected at once, the database could be switched from MySQL to Redis, a NoSQL database that provides real-time and high-efficiency services for data storage, increasing the I/O speed by storing all key-value data in the memory.

These, thought, would introduce a cost for the AWS services, but offering high efficiency and modularity.\\
		

\section{Conclusions}\label{conclusions}
Air pollution is an increasing problem requiring immediate attention.
All around the world, shared methods of transportation
have started growing in popularity and the more people use them, the better air quality gets in terms of lower quantities of pollution released in the air.

Centralized solutions are currently the most used to detect air pollution particles, but they only collect data in their proximity.
This paper describes the implementation of a proof of concept wireless IoT device, ArduECO, capable of gathering data from air, analyze and display them to the user (both citizens and municipalities).

ArduECO uses an Arduino-like board (NodeMCU) with a built-in Wi-Fi to send data recorded from an MQ-7 sensor for CO and a GPS module, in the cloud.
These gathered, aggregated and elaborated data can then be displayed on a frontend map, with each marker indicating a reading from the device.

At this first stage, ArduECO is currently intended to be integrated with bike sharing fleets but it can be easily and generally extended.
We have hence specifically focused on the idea of creating a pervasive sensing system, based on bikes, to gather, aggregate, elaborate and share data about air pollution in urban environments. To this aim, we have installed our prototype on a bicycle and conducted some preliminary tests.
From this preliminary work, plenty of improvements and future directions can be considered: quantity/quality/thrustworthiness of sensed data \cite{Dener15}, opportunistic networking of mobile devices\cite{Palazzi18}, data anonymity and privacy\cite{Furini20}, energy consumption\cite{Ciman14}, data analysis\cite{Delnevo20}, geosocial search\cite{sammarco17}, mobility models, city planning\cite{Chourabi11}, etc.
	
We are currently planning both to add more sensors on our solution and to perform a more extensive testing campaign.

\end{document}